\documentclass[12pt,preprint]{aastex}
\newcommand{\beq}{\begin{equation}}
\newcommand{\eeq}{\end{equation}}
\newcommand{\beqy}{\begin{eqnarray}}
\newcommand{\eeqy}{\end{eqnarray}}
\newcommand{\bce}{\begin{center}}
\newcommand{\ece}{\end{center}}

\shorttitle{Skyrmion Stars}
\shortauthors{P.~Jaikumar \& R.~Ouyed}
\begin{document}

\title{Skyrmion Stars\\
       Astrophysical motivations and implications}

\author{Prashanth Jaikumar}
\affil{Physics Division, Argonne National Laboratory, 9700 S. Cass Avenue,\\ Argonne, IL 60439 USA
       }
\email{jaikumar@phy.anl.gov}

\and

\author{Rachid Ouyed}
\affil{Department of Physics and Astronomy, University of Calgary,
       2500 University Drive NW, Calgary, AB T2N 1N4 Canada}
\email{ouyed@phas.ucalgary.ca}

\begin{abstract}
We study mass-radius relations for compact stars employing an equation of state (EOS) of dense matter based on a Skyrme fluid. The zero-temperature mean-field model is based on mesonic excitations, incorporates the scale breaking of QCD, and accommodates baryons (nucleons) which arise as a solitonic configuration of mesonic fields. Stable configurations are obtained for central densities $\rho_c/\rho_n\leq 5.0$ where $\rho_n=2.575\times 10^{14}$ g/cc is the nuclear saturation density. These {\it Skyrmion Stars} are mostly fluid, with a crust which we describe by the EOS of Baym, Pethick and Sutherland. Their masses and radii
 are in the range $0.4 \le M/M_{\odot} \le 3.6$ and $13~{\rm km} \le R \le 23~{\rm km}$, respectively.  The minimum spin period is computed to be between
 $0.7~{\rm ms}\le P \le 2.1~{\rm ms}$. They appear to have a mass-radius curve quite different from either neutron or quark stars, and provide a suitable description of the heavier mass neutron stars discovered recently due to the inherently stiff EOS. Within the same model, we compute the dominant neutrino emissivity in neutron-rich $\beta$-equilibrated matter, and determine the cooling behaviour of Skyrmion stars. 
\end{abstract}


\section{Introduction}
An oft-stated goal in the study of compact stars (neutron stars or
possibly even quark stars) is the determination of the phase of matter
in their interior through astronomical observations (Weber 1999,
Lattimer 2000, Heiselberg \& Pandharipande 2000).  An equation of
state is used to describe this phase; since it is an input into the
equations that describe the structural properties of the star, it can
therefore be constrained by stellar mass and radius observations (van
Kerkwijk 2001, Lattimer \& Prakash 2001, Lattimer \& Prakash
2004). The challenge in this field arises, on the one hand, from our
lack of theoretical knowledge about the EOS of matter at the highest
densities, and on the other hand, from mass and particularly radius
determination of compact stars (Haensel 2003, Heiselberg
2001). Despite improved many-body physics of nuclear and neutron
matter of relevance to neutron star interiors (Schwenk 2004, Schwenk
2003), and increasingly accurate estimations of mass and radius
(Manchester 2004), a number of EOS remain in use, with no definite or
universal answer as to which is the best one at supranuclear
densities. That being said, theoretical studies of the nuclear
equation of state can, to an extent, be gauged for their relative
success when confronted against data (particularly, accurate radius
measurements), and we refer the reader to the article by Lattimer \&
Prakash 2001 (and references therein) for details.

\vskip 0.2cm

\noindent Until recently, timing measurements of radio binary pulsars
yielded accurately measured neutron star masses in the range 1.26 to
1.45 $M_{\odot}$ (Thorsett and Chakrabarty 1999). Newer observations
of binary systems of pulsars and white dwarfs allow for a larger range
of masses; for example, the binary component PSR J0751+1807 with
2.1$_{-0.5}^{+0.4}$M$_\odot$ at 95\% confidence (Nice et al, 2005). We
also point out the possible confirmation (at 95\% confidence) of at
least one pulsar with $M>1.68M_{\odot}$ by Ransom et al, (2005),
following the recent detection of 21 millisecond pulsars in the
globular cluster Terzan 5, 13 of which are in binaries. These
observations of presumably heavier mass neutron stars seem to favor a
stiffer equation of state than has been explored in the literature, at
least for the central densities that we consider in this work
($\rho_c\leq 5.0\rho_n$). Intriguingly, it has been shown in Ouyed \&
Butler (OB98) that an equation of state based on the solitonic Skyrme
model yields quite stiff equations of state and neutron star masses
greater than 2$M _{\odot}$ for standard radii of 15 km or so (while
still obeying causality limits). In this letter, we extend the work
done in OB98 in two main directions. We construct the equation of
state for the Skyrme fluid including isospin breaking effects in order
to address asymmetric matter. This is more realistic than the approach
in OB98, where only symmetric matter and pure neutron matter was
considered, since neutrons begin to drip out of nuclei already at
sub-nuclear densities, and the neutron star is generally characterized
by an asymmetric mixture. We contrast the resulting mass-radius curve
with that of neutron and quark stars constructed out of more
conventional equations of state. We also address neutrino cooling
within the same model to obtain a more complete picture of the
behaviour of Skyrmion stars, and compare our cooling curves to
temperature data. Our main goal in this paper is to motivate, analyze
and elucidate the consequences of using the Skyrme model as a
description of compact star interiors, particularly in light of the
recent discovery of neutron stars with masses greater than the
canonical 1.4$M_{\odot}$. This paper is organized as follows: In \S 2,
we consider observational hints for large mass neutron stars. In \S 3,
we describe the Skyrme model, its extension to dense matter at nuclear
densities and beyond, and obtain an equation of state. In \S 4, we
display the mass-radius curves that are obtained from this EOS, and
consider dominant channels for neutrino emission and cooling of the
star. Our findings and conclusions are detailed in \S 5. The
Mathematica routines used in constructing the EOS for our model, as
well as those used to generate cooling curves, can be accessed on the
public domain {\it http://www.capca.ucalgary.ca} and downloaded for
personal use.

\section{Observational motivation}

We are closing in on neutron stars both observationally (masses and
radii) and theoretically (EOS at high density). Observations of 
X-ray binaries offer the best support for larger masses with values of
$1.8-2.4M_{\odot}$ (Quaintrell et al, 2003), $2.4\pm 0.27M_{\odot}$
(Clark et al, 2002) and $2.0-4.3M_{\odot}$ (Shahbaz et al, 2004) being
reported. Although these measurements are not as precise as for radio
pulsars~\footnote{We should note that there is only one NS in a
binary radio pulsar system, the pulsar J0751+1807 for which 
a large mass $2.1\pm 0.5M_{\odot}$ has been derived (Nice et al. 2004).}
they seem to hint at the existence of heavy neutron stars.
\vskip 0.2cm
\noindent
Such massive NSs are further suggested indirectly by (i) a possible
 interpretation, within current uncertainties in age and surface
 temperatures, of some colder neutron stars such as Vela and Geminga
 pulsars as moderately heavy stars ($\sim 1.5-1.8M_{\odot}$). These
 suggestive calculations are based on models of neutrino cooling
 taking proton and neutron superfluidity into account (Kaminker et
 al. 2001, Gusakov et al. 2005). It should be noted, however, that an
 interpretation of these pulsars as low mass pulsars, being consistent
 within the same uncertainties, cannot be ruled out (Page et
 al. 2004).  (ii) modeling of neutrino-driven winds from very massive
 ($M\sim 2.0M_{\odot}$) and compact ($R\sim 10$ km) neutron stars
 which reproduced the r-process abundance pattern with reasonable
 success (Otsuki et al, 2000; Sumiyoshi et al, 2000), due to the short
 expansion timescale and high entropy that are required to obtain a
 high neutron-to-seed ratio. However, such a large mass can be
 criticized from the preponderance of observational data on neutron
 star mass; (iii) recent studies by Popov \& Prokhorov (2005)
 addressing the issue of existence and formation of massive Skyrmion
 stars.  They considered different channels for the formation of
 massive rapidly rotating neutron stars using a population synthesis
 code to estimate numbers of massive neutron stars on different
 evolutionary stages, and found that a significant increase in neutron
 star mass due to accretion is possible for certain values of initial
 parameters of the binary. They concluded that, within the Skyrmion
 star scenario, rotating neutron stars with $M\geq 2M_{\odot}$ can
 arise from binary evolution. The authors argued that a significant
 part of such heavy stars can be observed as millisecond radio
 pulsars, as X-ray sources in pair with white dwarfs, and as accreting
 neutron stars with very low magnetic fields.  These massive Skyrmion
 star candidates (with the above mentioned observable signatures) can
 hopefully be tested in the near future; we refer the interested
 reader to Popov \& Prokhorov (2005) for more details.
\vskip 0.2cm
\noindent The evidence for large neutron star masses is still being debated
 and one awaits confirmation. However,
 the large masses hinted at above pose a challenge to models of neutron stars
based upon modern EOS even as uncertainties are being reduced by improved
two- and three-body forces, relativistic effects
and many-body calculations (Heiselberg \& Pandharipande 2000).
Even the stiffest EOS developed so far seems to be facing
difficulty in accounting for the extreme values
(up to $2.4M_{\odot}$; see discussion in
Heiselberg 2001). This brings us to a theoretical
discussion of the Skyrme model as an alternative.

\section{Theoretical motivation}
\subsection{The Skyrme model}
Well before the discovery of Quantum Chromodynamics (QCD) as the fundamental theory of the strong force, Skyrme (1962) provided a model based on pionic excitations, in which the nucleon arises as a classical configuration of the pion fields with a non-trivial topology and a localized energy density (a soliton). This is remarkable in that it allows us to consider fermions (nucleons) in a bosonic (pionic) theory without introducing any explicit fermion fields. The Skyrme model has been placed on a sound theoretical footing in subsequent analyses (Witten 1979) and it is now accepted as a valid description for certain aspects of the strong force. Its connection to phenomenological nuclear forces is well established (Nyman and Riska 1990), and its justification based on large $N_c$ arguments ('t Hooft 1974) stemming from a quark-gluon picture indicates that it acts as a ``bridge'' between QCD and nucleon-nucleon ($n-n$) forces. Several detailed reviews on the Skyrme model may be found in the literature (Holzwarth and Schwesinger 1986, Zahed and Brown 1986, Schechter and Weigel 2000). 

\subsection{The Skyrme model and gravity}

Skyrmions coupled to gravity have been previously explored in Glendenning et al. 1988, Droz et al. 1991, Bizon and Chmaj 1992, with the conclusion that Skyrmions with large topological charge $N$ cannot be models of neutron stars due to gravitational collapse at micro-sizes ($N\sim 10^{18}$) compared to neutron stars ($N\sim 10^{57}$). Furthermore, they are unstable to $N=1$ configurations if non-spherical shapes are disallowed. Therefore, we adopt a more conservative approach in which the star is composed of an interacting ensemble (at the mean-field level) of $N=1$ skyrmions (unit baryon number). In the spirit of nuclear mean-field approaches to the equation of state, we utilize the skyrme model for the microscopics, and couple the ensemble of $N=1$ skyrmions to gravity through the stress-energy tensor $T^{\mu\nu}$ of the dilute skyrme fluid (no overlap between skyrmions). This implies that gravity becomes strong only at very large $N$ (unlike in afore-mentioned studies where the entire star is treated as a topological object coupled to gravity) and that instability towards single-particle emission is not relevant in our model. The skyrmion profile and its mass-radius characteristics are very different than obtained in previous works. This approach will provide for a  much more realistic description of neutron stars, while maintaining a connection to the Skyrme model. 
 
\subsection{The Skyrme fluid}

For astrophysical applications, one needs a many-body description of the nuclear force, and for neutron star densities, an in-medium approach based on the relevant degrees of freedom in the theory. As a successful and theoretically sound model, the Skyrme Lagrangian bears further investigation as regards its application to astrophysics. A particularly useful mean-field model was developed in the work of K\"{a}lbermann (1997), who studied Skyrmions in a dilute fluid approximation, where auxiliary dilaton ($\sigma$) and vector ($\omega$) fields couple to the Skyrmion and carry the information of density and temperature. The (non-zero) expectation values of these background fields are determined from the Skyrme Lagrangian itself, in the spirit of the mean field approach. 

\vskip 0.2cm

\noindent The Lagrangian for the Skyrme model, augmented by the $\sigma,\omega$ fields, and including isospin-breaking effects from the $\rho$ meson as well as explicit scale-breaking effects from the dilaton and quark masses is given by
\begin{eqnarray}
{\cal L}&=&{\cal L}_2 + {\cal L}_4 + {\cal L}_{m_{\pi}}- V_B - V_{\sigma} + {\cal L}_{\omega}+{\cal L}_{\rho}-V_{I} \label{Lsky}\\ \nonumber 
&=&{\rm e}^{2\sigma}\left[\frac{1}{2}\Gamma_0^2\partial_{\mu}\sigma\partial^{\mu}\sigma - \frac{F_{\pi}^2}{16}{\rm Tr}(L_{\mu}L^{\mu})\right]+\frac{1}{32{\rm e}_s^2}{\rm Tr}[L_{\mu},L_{\nu}]^2\\ \nonumber
 &+&{\rm e}^{3\sigma}\frac{F_{\pi}^2m_{\pi}^2}{8}{\rm Tr}(U-1) - g_{\rm V}\omega_{\mu}B^{\mu} - B[1+{\rm e}^{4\sigma}(4\sigma-1)]\\ \nonumber
 &-& \frac{1}{4}(\partial_{\mu}\omega_{\nu}-\partial_{\nu}\omega_{\mu})^2 + \frac{1}{2}{\rm e}^{2\sigma}m_{\omega}^2\omega_{\mu}\omega^{\mu}-\frac{1}{4}F_{\mu\nu}^{\rho}F^{\rho,\mu\nu}+\frac{1}{2}{\rm e}^{2\sigma}m_{\rho}^2\vec{\rho}_{\mu}\vec{\rho}^{\mu}-g_{\rho}\vec{\rho}_{\mu}\vec{I}^{\mu} \,\,, 
\end{eqnarray}
where 
\beq
L_{\mu}=U^{\dag}\partial_{\mu}U\quad.
\eeq
In eqn.(\ref{Lsky}), $F_{\pi}$ is the pion decay constant, ${\rm e}_s$ is the Skyrme parameter, and $g_{\rm V}$ is the $\omega~N$ coupling strength. The $\omega$-meson couples to the topological (baryon) current

\beq
B^{\mu}=\frac{\epsilon^{\mu\alpha\beta\gamma}}{24\pi^2}{\rm Tr}[(U^{\dag}\partial_{\alpha}U)(U^{\dag}\partial_{\beta}U)(U^{\dag}\partial_{\gamma}U)]\quad.
\eeq

The $\rho$ meson, introduced here to extend our model to asymmetric
matter, couples to the isospin current $\vec{I}^{\mu}$ with coupling
$g_{\rho}$, and its field strength is non-abelian
$F_{\mu\nu}^{\rho}=\partial_{\mu}\vec{\rho}_{\nu}-\partial_{\nu}\vec{\rho}_{\mu}+g_{\rho}(\vec{\rho}_{\mu}\times\vec{\rho}_{\nu})$. The
$\rho$ meson is viewed here as a massive Yang-Mills field, and the
model is renormalizable with the introduction of a Higgs field. To
obtain the equation of state, we will work in the limit of infinite
Higgs mass, which simplifies the equations of motion for the mean
fields (Serot and Walecka 1986). This is a standard approach in
$\rho,\omega,\sigma$-type Walecka models (Quantum hadrodynamics) which
closely resembles our model. Scale invariance is broken by quantum
fluctuations of the QCD vacuum and finite quark masses, and this
feature is manifest since the dilaton mocks up the scale anomaly of
QCD (Nielsen 1977, Collins et al, 1977) \beq
T_{\mu}^{\mu}=\partial_{\mu}D^{\mu}=-\frac{9\alpha_s}{8\pi}G^{a}_{\mu\nu}G^{a\mu\nu}+\sum_qm_q\langle\bar{q}q\rangle\quad ,
\eeq where $T_{\mu}^{\mu}$ is the trace of the energy-momentum tensor,
$D^{\mu}=T^{\mu\nu}x_{\nu}$ is the dilatation current, $\alpha_s$ the
strong coupling constant, and $G^{a}_{\mu\nu}$ the gluon field
strength. The sum over $q$ runs over the up and down quark flavors,
and we have ignored the anomalous dimension. Consistency with the
trace anomaly demands that the quark condensate behaves as
$\langle\bar{q}q\rangle\rightarrow\langle\bar{q}q\rangle{\rm
e}^{3\sigma}$ under a scale transformation $r\rightarrow r{\rm
e}^{-\sigma}$. Its connection to eqn.(\ref{Lsky}) is through the
Gell-Mann Oakes Renner relation

\beq
F_{\pi}^2m_{\pi}^2=-2(m_u+m_d)\langle\bar{q}q\rangle \quad.
\eeq

\noindent From the above Lagrangian, it is easily verified that the gluonic contribution 

\beq 
T_{\mu}^{\mu}=(\Gamma_{0}{\rm e}^{\sigma})^4\quad.
\eeq 
In more familiar terms, $\Gamma_0^4/{\rm e}=-4B$, where $B$ is the bag
constant. The dilaton field also provides the required
intermediate-range attraction in the $NN$ force.  The $\omega$-meson
provides the right sign for the isospin-independent spin-orbit
force. It can also stabilize the soliton in the absence of the Derrick
term (${\cal L}_4$)(Adkins and Nappi 1984a). For the sake of
completeness, we have included the heretofore ignored pion mass term,
which has recently shown to be important for multi-skyrmion systems
(Battye and Sutcliffe 2005). In fact, these authors find that
alternate minima of the Skyrme energy functional can arise if either
the baryon number or the pion mass is large. For the (physical) value
of the pion mass, and the product ansatz for the Skyrmions that we use
in this work, we can continue to use the ``hedgehog''
ansatz~\footnote{\noindent The product ansatz is used for simplicity. For truly
multi-skyrmion shell-like solutions with large baryon numbers of 20 or
more, the pion mass plays an important role.}  
\beq
 U={\rm exp}(i\vec{\bf \tau}.\hat{\bf r}F(r))\,\,, 
\eeq 
with ${\bf \vec{\tau}}$ the Pauli matrices. With some algebra, one can show that the Skyrmion mass is 

\beqy 
M&=&M_0{\rm e}^{\sigma_0}+M_{\pi N};\quad
M_0=4\pi\int~dr~r^2M_0(r)\,;\nonumber\\M_0(r)&=&\frac{F_{\pi}^2}{8}\left[{F^{\prime}}^2+\frac{2{\rm
sin}^2F}{r^2}\right]+\frac{{\rm sin}^2F}{2{\rm e}_s^2r^2}\left[\frac{{\rm
sin}^2F}{r^2}+2{F^{\prime}}^2\right]\,\,,\\ 
M_{\pi N}&=&\pi m_{\pi}^2F_{\pi}^2\int~dr~r^2(1-{\rm cos} F)\,\,,
\label{eq:mstar}
\eeqy

where $\sigma_0=\langle\sigma\rangle$, $\omega_0=\langle\omega\rangle$ and $\rho_0=\langle\rho\rangle$ henceforth denote mean-field values. The resemblance of the above form to the free Skyrmion case comes from the association of $\sigma$ with broken scale invariance. With the product ansatz for the Skyrmion

\beq
U_{B=N}({\bf r,R_1,..,R_N})=U({\bf r-R_1})..U({\bf r-R_N})\quad ,
\eeq

and the sum ansatz for the $\sigma,\omega,\rho$ fields

\beq
\sigma_{B=N}=\sigma_1+..+\sigma_N;\quad \omega_N=\omega_1+..+\omega_N;\quad \rho_N=\rho_1+..+\rho_N\,\,,
\eeq

the mean-field treatment implies that the equations for the Skyrme fluid may be derived by solving the Euler-Lagrange equations from ${\cal L}$ for a single free Skyrmion in the absence of background fields, and rescaling as

\beq
r\rightarrow{\rm e}^{-\sigma_0}r,\quad \omega\rightarrow {\rm e}^{\sigma_0}\omega,\quad \rho\rightarrow {\rm e}^{\sigma_0}\rho
\eeq
in the presence of the medium (K\"albermann 1997). The free Skyrmion mass $M_0$  and the contribution from explicit scale breaking (the pion-nucleon sigma term $M_{\pi N}$) can be formally obtained by noting that the solution for $F(r)$ is (Adkins and Nappi 1984b)
\beq
F(r)=0.759\frac{{\rm e}^{-m_{\pi}r}}{r}\quad.
\eeq
The ``effective mass'' of the Skyrmion in the fluid is then $M$. We now have the tools to construct the equation of state for the Skyrme fluid.
\subsection{Equation of State}

To obtain the equation of state, we require a relation between the pressure and energy density of the fluid. In the mean field approximation, the energy of $N$ Skyrmions per unit volume is given by
\beqy
E_V&=&2\int\frac{d^3p}{(2\pi)^3}\left[E_p^{\rm prot}(n_p+\bar{n}_p)^{\rm prot}+E_p^{\rm neut}(n_p+\bar{n}_p)^{\rm neut}\right] + V_{\sigma}(\sigma_0) \nonumber\\
&-&\frac{1}{2}{\rm e}^{2\sigma_0}m_{\omega}^2\omega_0^2 + g_V\omega_0\rho_V\nonumber\\
&-&\frac{1}{2}{\rm e}^{2\sigma_0}m_{\rho}^2\rho_0^2 + \frac{1}{2}g_{\rho}\rho_0\rho_I\nonumber\\\label{mfenergy}
\eeqy
where\footnote{the extra 1/2 factor in the $\rho$ coupling compared to $\omega$ coupling is due to the conventional definition of the Pauli isospin matrices} $\rho_V$ is the net (baryon minus antibaryon) density and $\rho_I$ is the net (proton minus neutron) isospin density. We choose $m_{\rho}=770$ MeV, and we will fit the binding energy for symmetric matter (-16 MeV per nucleon) and symmetry energy (32 MeV) at saturation to obtain the parameters $g_V$ and $g_{\rho}$ respectively. Here, the particle and anti-particle distribution functions are

\beq
n_p=\left({\rm exp}\frac{\epsilon_p-\mu}{T}+1\right)^{-1};~ \bar{n}_p=\left({\rm exp}\frac{\bar{\epsilon}_p+\mu}{T}+1\right)^{-1}
\eeq

with the dispersion relations $\epsilon_p^{\rm prot}=E_p^{\rm prot}+g_V\omega_0+g_\rho\rho_0/2$, $\epsilon_p^{\rm neut}=E_p^{\rm neut}+g_V\omega_0-g_\rho\rho_0/2$, $\bar{\epsilon}_p^{\rm prot}=E_p^{\rm prot}-g_V\omega_0+g_{\rho}\rho_0/2$, $\bar{\epsilon}_p^{\rm neut}=E_p^{\rm neut}-g_V\omega_0-g_{\rho}\rho_0/2$ where $E_p^{\rm prot}=E_p^{\rm neut}=\sqrt{p^2+M^2}$ (our model yields the same Dirac effective mass for neutrons and protons). The chemical potentials satisfy

\beq
\epsilon_p^i(p_{F_i})=\mu_i\,;\quad i={\rm prot, neut}
\eeq

\noindent where $p_{F_i}$ is the Fermi momentum of species $i$. The quantity $E_p$ is related to the energy of an individual Skyrmion $E_{sk}$ by a Lorentz boost of the static solution (K\"{a}lbermann 1997). It implicitly includes a contribution from the pion mass term from Eq.(\ref{eq:mstar}).

\noindent Since the scale of thermal excitations in the Skyrme fluid is much less than the baryon chemical potential, we will work at $T=0$ for this section.~\footnote{While considering neutrino cooling, we naturally admit finite temperatures.} The Fermi distributions reduce to theta functions, and we find

\beq
\rho_V=\frac{\left(p_{F_{\rm prot}}^3+p_{F_{\rm neut}}^3\right)}{3\pi^2}\,;\quad\rho_I=\frac{(p_{F_{\rm prot}}^3-p_{F_{\rm neut}}^3)}{3\pi^2}\quad .
\eeq

\noindent The equations of motion for the mean fields are obtained by minimization of the energy eqn.(\ref{mfenergy}) with respect to the $\sigma$, $\omega$ and $\rho$ fields :

\beqy 
\label{mfeom}
0&=&\frac{\partial
E_V}{\partial\sigma_0}=\frac{2}{(2\pi)^3}\left(\int^{p_{F_{\rm
prot}}}d^3p+\int^{p_{F_{\rm neut}}}d^3p\right)\left(\frac{\partial
E_p}{\partial\sigma_0}\right)+\frac{dV(\sigma_0)}{d\sigma_0}\nonumber\\
&-&{\rm e}^{2\sigma_0}m_{\omega}^2\omega_0^2-{\rm
e}^{2\sigma_0}m_{\rho}^2\rho_0^2\quad .\\ 0&=&\frac{\partial
E_V}{\partial\omega_0}={-\rm
e}^{2\sigma_0}m_{\omega}^2\omega_0+g_V\rho_V\quad .\\
0&=&\frac{\partial E_V}{\partial\rho_0}={-\rm
e}^{2\sigma_0}m_{\rho}^2\rho_0+\frac{g_{\rho}}{2}\rho_I\quad .\\\nonumber
\eeqy 

\noindent The symmetry energy can be calculated in our model to be

\beq
a_s(\rho_V,\sigma_0(\rho_V))=\frac{1}{2}\rho_V\left(\frac{\partial^2E_V}{\partial\rho_I^2}\right)_{\rho_I=0}=\frac{g_{\rho}^2p_F^3}{12\pi^2m_{\rho}^2}+\frac{1}{6}\frac{p_F^2}{\sqrt{p_F^2+M^2}}\quad ,
\eeq

\noindent where $p_F=(3\pi^2\rho_V/2)^{1/3}$. At saturation $\rho_V=\rho_n=0.154$ fm$^{-3}$ and with an effective mass of $M=600$ MeV and a symmetry energy of 32 MeV~\footnote{The symmetry energy in dense matter determines the threshold for the direct urca process involving protons ($n\rightarrow p+e^-+\bar{\nu}_e$). This is an extremely efficient neutrino cooling mechanism, and will be addressed in the section on cooling.}, we get $g_{\rho}=7.70$. In asymmetric matter, beta-equilibrium demands $\mu_{\rm neut}=\mu_{\rm prot}+\mu_{\rm elec}$, while local charge neutrality dictates that $n_{\rm prot}=n_{\rm elec}$. These conditions can be substituted back in the mean field equations as follows :\\ 
Let

\beq
x=\frac{p_{F_{\rm neut}}}{p_{F_{\rm
prot}}};\quad y=\frac{n_{\rm prot}}{n_{\rm neut}+n_{\rm prot}}=\frac{1}{x^3+1}\,,
\eeq

\noindent The energy $E_V$ can be expanded about the symmetric point $y=1/2$ to a good approximation as

\beq
E_V(\rho_V,y)=E_V(\rho_V,1/2)+a_s(\rho_V,\sigma_0(\rho_V))(1-2y)^2+...\quad ,
\eeq

\noindent where ... are ignorable corrections. Studies of neutron matter show that this expansion works well over a wide range of values for $y$ and $\rho_V$ (Prakash et al. 1988).  Then, beta equilibrium along with charge neutrality implies

\beq
(3\pi^2\rho_V y)^{1/3}=4a_s(\rho_V,\sigma_0(\rho_V))(1-2y)\quad ,
\eeq

\noindent which fixes a solution $y=y_0$ for any baryon density. It follows that

\beq
{p_{F_{\rm prot}}}=(3\pi^2\rho_V y_0)^{1/3}\,;\quad {p_{F_{\rm neut}}}=(3\pi^2\rho_V (1-y_0))^{1/3}\,;\quad \rho_I=\rho_V(2y_0-1)\,.\label{symeq}
\eeq

\noindent Performing the momentum integrations in eqn~(\ref{mfeom}) analytically, the equations of motion become

\beqy
\label{vsigma}
0&=&\frac{M^2}{4}\biggl[p_{F_{\rm prot}}\sqrt{{p_{F_{\rm prot}}}^2+M^2}+p_{F_{\rm neut}}\sqrt{{p_{F_{\rm neut}}}^2+M^2} \nonumber\\
&-&\frac{M^2}{2}{\rm ln}\biggl(\frac{[p_{F_{\rm prot}}+\sqrt{{p_{F_{\rm prot}}}^2+M^2}][p_{F_{\rm neut}}+\sqrt{{p_{F_{\rm neut}}}^2+M^2}]}{M^2}\biggr)\biggr]\nonumber\\
&+&\frac{dV(\sigma_0)}{d\sigma_0}-{\rm e}^{2\sigma_0}m_{\omega}^2\omega_0^2-{\rm
e}^{2\sigma_0}m_{\rho}^2\rho_0^2\quad . \\ 
0&=&\frac{\partial
E_V}{\partial\omega_0}={-\rm
e}^{2\sigma_0}m_{\omega}^2\omega_0+g_V\rho_V\quad .\\
0&=&\frac{\partial E_V}{\partial\rho_0}={-\rm
e}^{2\sigma_0}m_{\rho}^2\rho_0-\frac{g_{\rho}}{2}\rho_V(1-2y_0)\quad .\\\nonumber
\eeqy 

\noindent For symmetric matter, a fit to the binding energy $B_V=E_V/\rho_V-M$ of -16 MeV at saturation yields $g_V=7.54$ (OB98). In our case, for neutron-rich matter ($y\ll 1$), this value leads to a positive $B_V$ (see Figure~\ref{fig:binding}), implying that the matter is unbound, as expected. We do not expect $g_V$ to be sensitive to the isospin composition, as the $\omega$-meson couples to the net baryon density. Adopting this value of $g_V$, the above equations can be solved for the mean fields to yield the energy per baryon from eqn~(\ref{mfenergy}) at any baryon density.

\noindent The pressure of this ensemble is given by

\beq
P_V=\rho_V^2\frac{\partial(E_V/\rho_V)}{\partial\rho_V}\quad.
\eeq

\noindent The respective equations of motion show that the contribution of the vector meson $\omega$ to the pressure grows with density and is positive, a consequence of acting in the repulsive channel. The dilaton $\sigma$ gives a negative pressure and tends to decrease the (magnitude of) binding energy. The $\rho$ meson acts to stiffen the equation of state, particularly at higher densities. These features drive the pressure-density and mass-radius relations that we obtain in the following section.

\subsection{Fit parameters for neutron-rich matter}

In order to reproduce neutron matter phenomenology, four free
parameters ($a_i$, $i=1,4$) are added to the conventional dilaton
potential.\footnote{There is also the
possibility of modifying the $\omega$ field potential (see
K\"albermann (1997) and references therein).} Then,

\begin{eqnarray}
V_{\sigma} = &B&[1+{\rm e}^{4\sigma}(4\sigma-1)]
+ B [a_1({\rm e}^{-\sigma}-1) +
a_2({\rm e}^{\sigma}-1)\nonumber \\ 
&+& a_3({\rm e}^{2\sigma}-1) + a_4({\rm e}^{3\sigma}-1)]\,\,,
\end{eqnarray}
with the ``bag constant'' $B\sim (240$~MeV)$^4$.
Terms of the form ${\rm e}^{n\sigma}$
are the only ones that can be added to the potential in
accordance with the anomaly condition
\begin{equation}
{dV_{\sigma}\over d\sigma } = 0
\end{equation}
at $\sigma = 0$, implying $a_1 = a_2 + 2a_3 + 3a_4$.\footnote{
One chooses the term multiplied by $a_1$ to
have a negative power of $\sigma$ in order to avoid the
introduction of a second minimum in the potential
for $\sigma\le 0$. The only sensible minimum then
remains the one at $\sigma =0$.}
The remaining three parameters of the dilaton potential
are constrained by demanding that at saturation
density ($\rho_n=0.154$ Baryons/fm$^3$), the effective mass $M/M_0=0.60$, the compressibility
$K=270$~MeV, and that eqn.(\ref{vsigma}) is satisfied.
On fitting to these properties, we find

\begin{eqnarray}
a_1&=&-29689.7,\quad a_2=-274771.0,\\
a_3&=&222839.0, \quad a_4 = -66865.6\,.
\end{eqnarray}

Although large in comparison to the fitting parameters obtained in OB98, 
there are large cancellations that lead to typical values $V(\sigma)\simeq 0.06B$ for the entire range of $\sigma$ (and hence density) explored in this work. 
\begin{figure}[htb]
\begin{center}
\leavevmode
\epsscale{.8}
\plotone{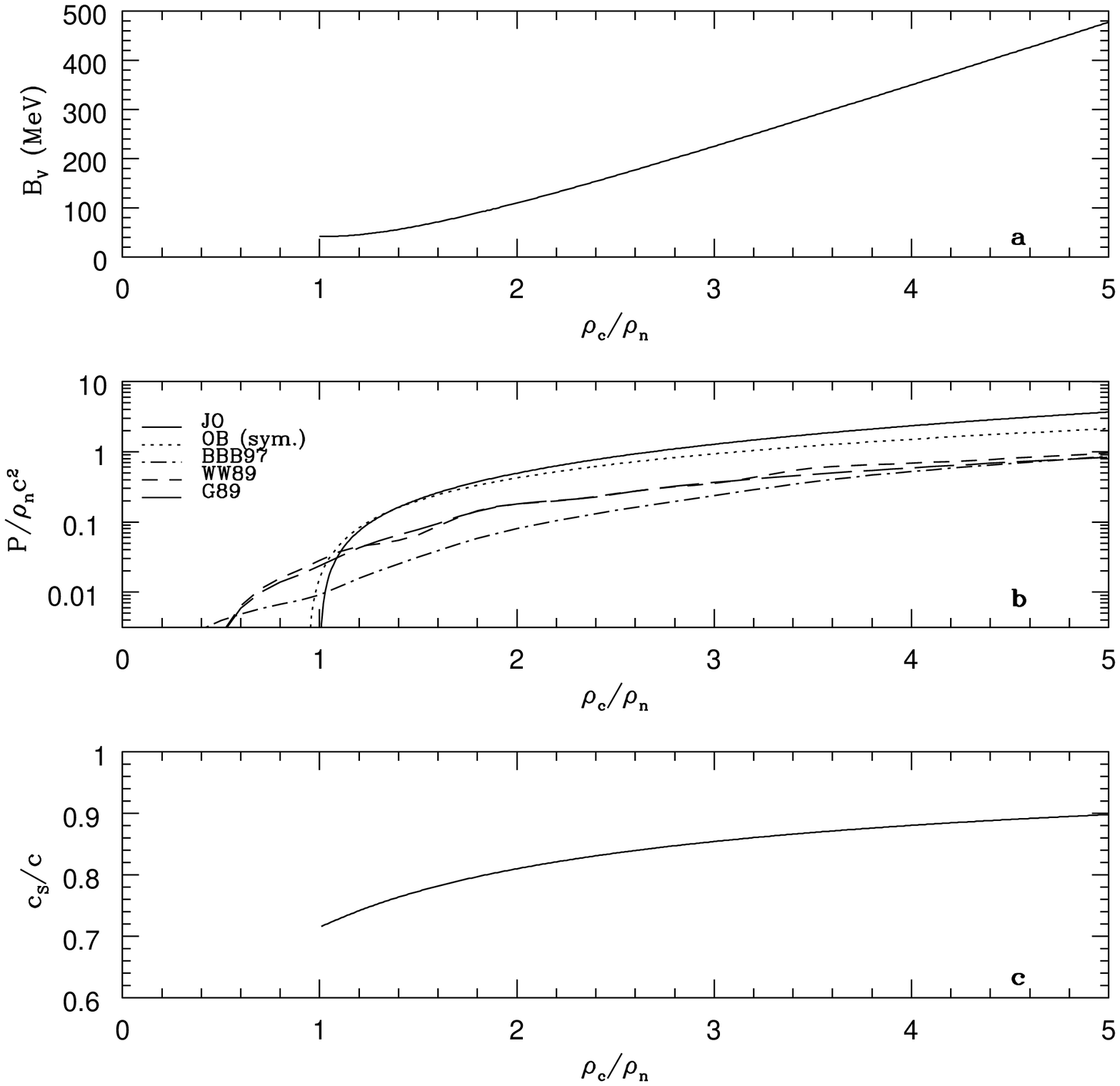}
\epsscale{1}
\end{center}
\caption{\emph{
{\bf a).} Binding energy per nucleon in neutron-rich matter, as a function
of the central density.  
{\bf b).} Pressure versus density for our model (JO) compared to OB98, BBB97 (Baldo et al. 1997), WW89 (Weber and Weigel 1989) and G89 (Glendenning 1989).
{\bf c).} Speed of sound (in units of $c$) for our model as a function of density.}}
\label{fig:binding}
\end{figure}
%
\noindent Figure~\ref{fig:binding} shows the binding energy per nucleon
 as a function of the density as well as the pressure-density relation (EOS)
 and the sound speed. A comparison to more conventional equations of state, as 
 well as the EOS of OB98 is presented therein. The increased stiffness with respect to the results of OB98 is evident. It is due to the limited attraction provided by the dilaton, and the extra repulsive effects from the $\rho$-meson. We now turn to applications of the Skyrme fluid EOS. In the next section, we discuss the global structure of Skyrmion stars and their cooling properties on the same footing. 

\section{Implications of the Skyrme EOS}

\subsection{Mass-radius relations}
\begin{figure}[htb]
\begin{center}
\epsscale{.8}
\plotone{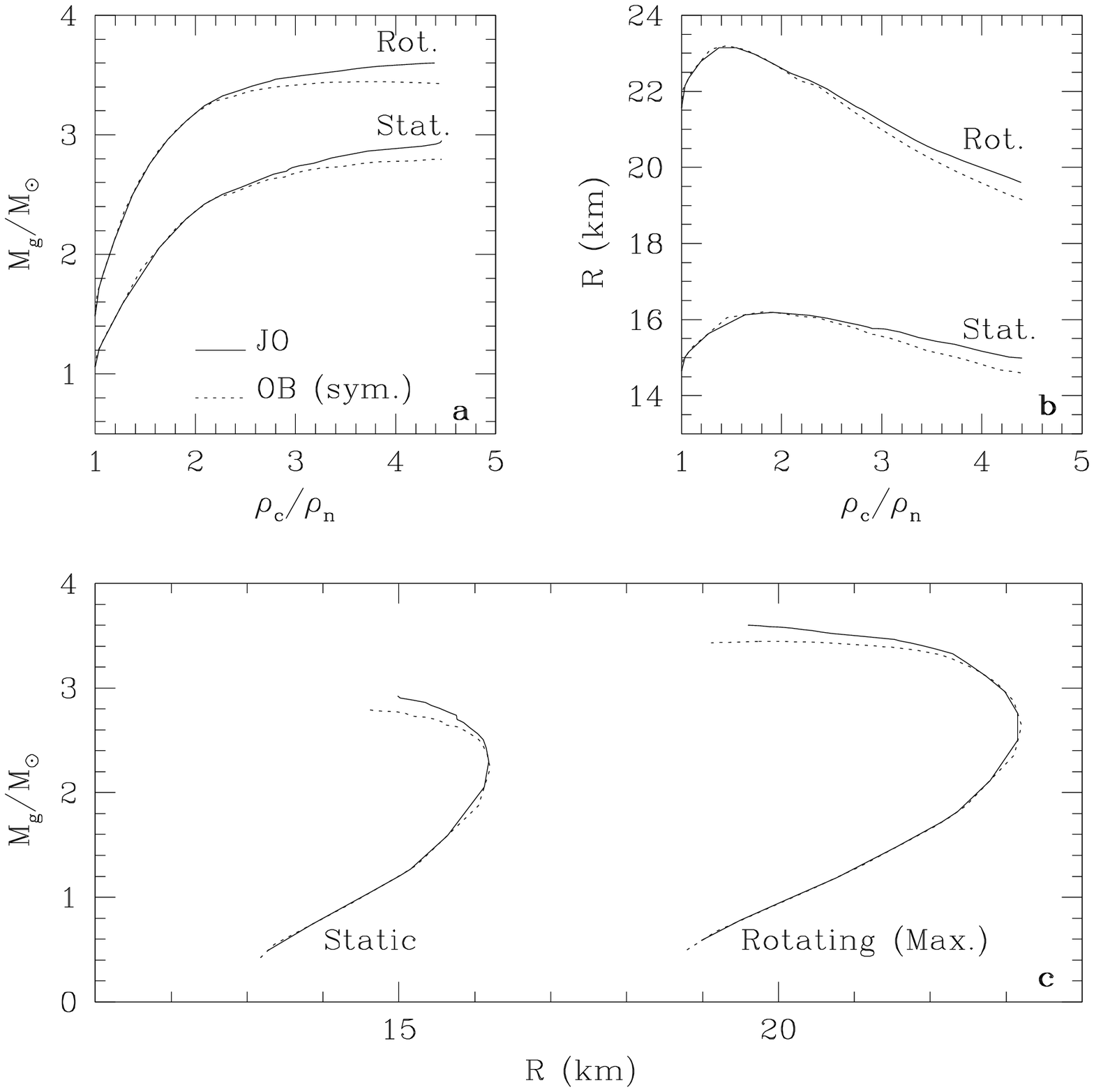}
\epsscale{1}
 \end{center}
\caption{\emph{
{\bf a).} Gravitational  Mass versus central density for zero temperature
Skyrmion stars in hydrostatic equilibrium. Both rotating and static configurations for our model and OB98 are shown to illustrate the differences. {\bf b).}
Radius versus central density for 
Skyrmion stars in hydrostatic equilibrium.
{\bf c).} The Mass$-$Radius plane.
}}
\label{fig:mvsr}
\end{figure}
\begin{figure}[htb]
\begin{center}
\epsscale{.8}
\plotone{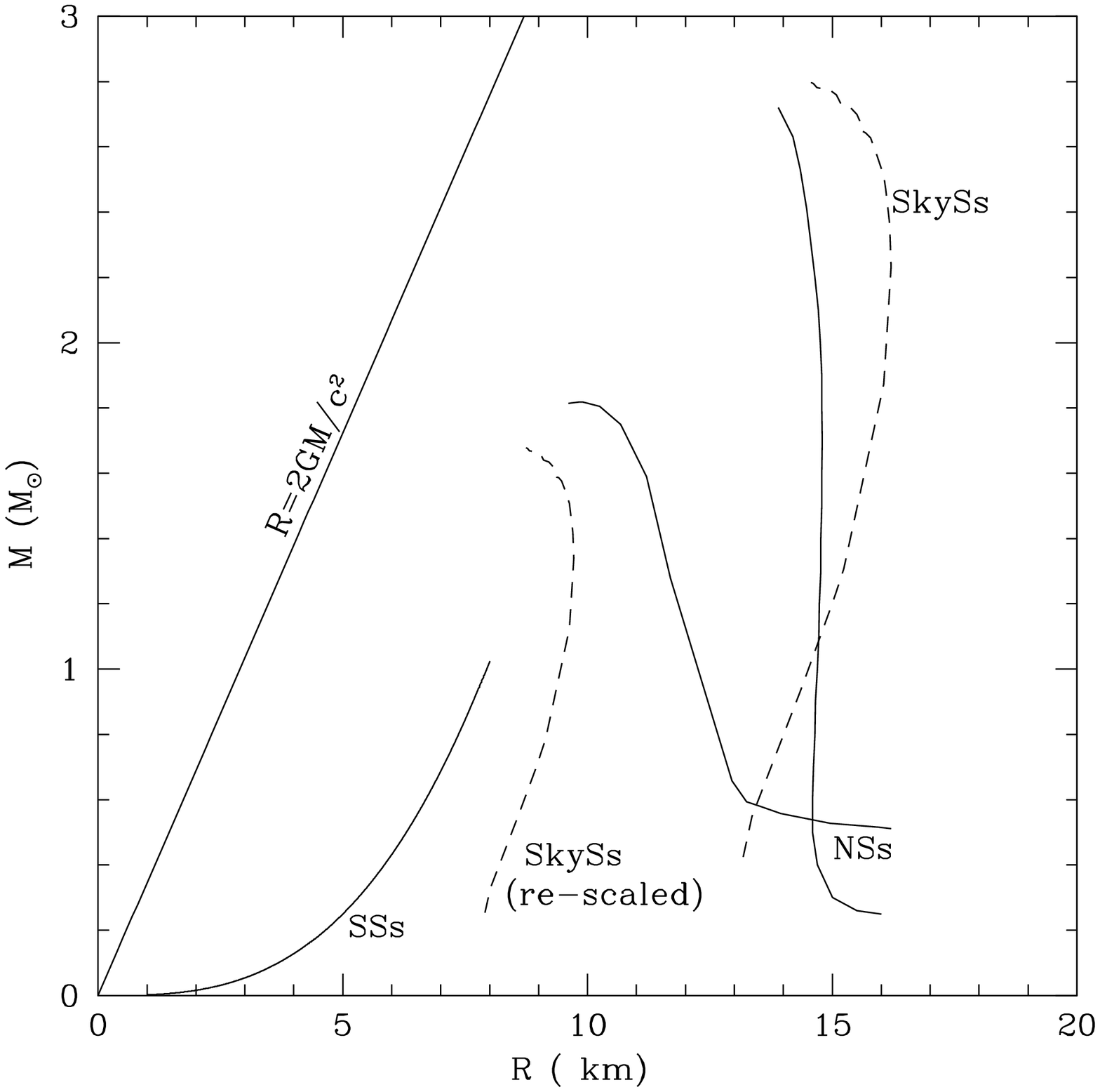}
\epsscale{1}
 \end{center}
\caption{\emph{Mass-Radius relationship characteristic
of neutron stars (NSs; e.g. Glendenning 1997 (soft), Mueller and Serot 1996 (stiff)) and quark stars (SSs;
e.g. Li et al. 1999) as compared to Skyrmion stars (SkySs). The SkySs 
model has been re-scaled for better comparison.
\label{fig:compare}
}}
\end{figure}
We proceed now to construct models for stars using the EOS developed
above. This is done by integrating the general-relativistic equation
of hydrostatic balance, including rotation, upto the surface of the
star where the pressure vanishes (Tolman 1934; Oppenheimer \& Volkoff
1939). For the rotating configurations we use the computer code {\it
RNS} (Stergioulas \& Friedman 1995; 1998).  We specify the equation of
state and the central energy density, and the code computes models
with increasing angular velocity until the star is spinning with the
same angular velocity as a particle orbiting the star at its equator
(see Ouyed 2002).

Figures \ref{fig:mvsr}a and \ref{fig:mvsr}b show the resulting stellar
masses and radii as a function of the central density, respectively.
For the rotating case, the masses are in the range $0.4\le
M/M_{\odot}\le 3.6$ while the radii (equatorial circumferential
radius; [proper equatorial circumference]/2$\pi$) are calculated to be
$18.6\ {\rm km}\leq R\leq 23.0\ {\rm km}$.  In Figure \ref{fig:mvsr}c,
we show the resulting Mass-Radius plane.  Note that the maximum mass
configuration does not correspond to the maximum radius case.  In
general, for a given central density, rotation allows the masses and
radii to increase by 30 \% and 40\%, respectively, when compared with
the non-rotating cases.  The amount of mass in the outer region of the
star (the crust region - $\rho < \rho_n$ - constructed using the
EOS of Baym, Pethick \& Sutherland, 1971) decreases with rotation.
The crust of rotating Skyrmion stars constitutes less than 5\% of the
total mass while it averages 20\% for the static configurations
(ref. Figure 5 of OB98).  The minimum spin period for Skyrmion stars using
 our improved EOS 
was computed to be $0.7\ {\rm ms}\le P \le 2.1\ {\rm ms}$ not very
different from what was found in Ouyed (2002).

\vskip 0.2cm

\noindent A comparison of the mass-radius curve for the Skyrmion star with that of neutron and quark stars (see Figure~\ref{fig:compare}) shows a qualitative difference that arises from the physics of the dilaton. The limited attractive force provided by the dilaton, and the inclusion of the $\rho$ meson leads to large stiffness even at moderate central pressures when compared to ordinary neutron stars. The radius of Skyrmion stars increases with mass within a certain mass window, despite increased gravitational forces, similar to the curve for self-bound quark stars (see rescaled curve in Fig. 3) and neutron stars described by a highly stiff equation of state (see curve labelled MS) based on a relativistic mean-field approach (eg. Mueller and Serot 1996).

\subsection{Neutrino cooling of the Skyrme fluid}
A unique method to probe the internal composition of a neutron star is
by tracing its temperature evolution with cooling simulations (Pethick
1992, Page 1998, Yakovlev et al. 2003). Neutron stars are born in
supernova explosions and as the neutron star cools, neutrinos begin to
free stream and essentially leave the star without further
interaction. As a consequence, about $30$ seconds after the explosion,
the cooling of neutron stars is controlled by neutrino
emission. Importantly, the neutron stars remain luminous enough during
the neutrino cooling era, so that the surface temperature can be
extracted from space telescope data and the theoretical cooling curves
can be confronted with observations. There is a host of well-known
neutrino emission processes that operate in the crust and core of the
neutron star (Yakovlev et al. 2001), and which one dominates the
cooling depends mainly on the temperature and to a lesser extent, on
the density. Since cooling rates are dependent on neutrino
emissivities, which in turn depend on the dense matter equation of
state employed, it is instructive to compute the dominant neutrino
emission rates. We now proceed to do so for the Skyrme fluid.  We
address only the global thermal energy balance of the star during the
neutrino and photon cooling eras, so that we may employ

\beq
C_V\frac{dT}{dt}=-L_{\nu}-L_{\gamma}  \label{cool}
\eeq
to compute the cooling history. The required inputs to the above
equation are the total specific heat at constant volume $C_V$, the
neutrino emissivity $\epsilon_{\nu}$ which determines the neutrino luminosity
as $L_{\nu}\sim \epsilon_{\nu}V$, where $V$ is the cooling volume of the star and $L_{\gamma}$, the photon luminosity which determines the late time ($>10^5-10^6$ yrs.) cooling behaviour. We will analyze this equation by approximating $L_{\gamma}=0$, which defines the neutrino cooling epoch, or $L_{\nu}=0$ which defines the photon cooling epoch. In asymmetric matter, the direct urca process (in medium $\beta$-decay) $n\rightarrow p+e^-+\bar{\nu}_e$ and $p+e\rightarrow n+\nu_e$ as well as the modified urca process $n+n\rightarrow n+p+e+\bar{\nu}_e$ and $n+p\rightarrow n+n+e+\nu_e$ become important channels of neutrino emission. The former is kinematically allowed only if energy and momentum conservation, expressed by the triangle inequality $\Theta(p_{F_e}+p_{F_p}-p_{F_n})$ is satisfied. This happens at a critical value of $y$. On using the above condition, the threshold where direct urca can start to occur is (Lattimer et al. 1991)

\beq
y_{\rm crit}=\frac{p_{F_p}^3}{(p_{F_p}+p_{F_e})^3+p_{F_p}^3}=\frac{1}{(1+\frac{p_{F_e}}{p_{F_p}})^3+1}\quad .
\eeq

Since charge neutrality demands $p_{F_e}=p_{F_p}$, this implies that 
\beq
y_{\rm crit}=\frac{1}{9}\quad .
\eeq

Only if $y$ exceeds 1/9 can we have a direct urca process. The associated neutrino rate is (Lattimer et al. 1991)

\beq
\epsilon_{\nu}^{\rm dUrca}=1.2\times 10^{27}\left(\frac{m_p^*}{\rm GeV}\right)\left(\frac{m_n^*}{\rm GeV}\right)\left(\frac{\rho_V}{\rho_n}\right)^{1/3}~y^{1/3}~\Theta(y-y_{\rm crit})~T_9^6~ {\rm erg~cm}^{-3}{\rm s}^{-1}\,.
\eeq

Here, the neutron and proton Landau effective masses~\footnote{The Landau effective masses $m_p^*$ and $m_n^*$ account for momentum-dependent interactions at the Fermi surface.} are a function of density and are given by

\begin{figure}[htb]
\begin{center}
\epsscale{.8}
\plotone{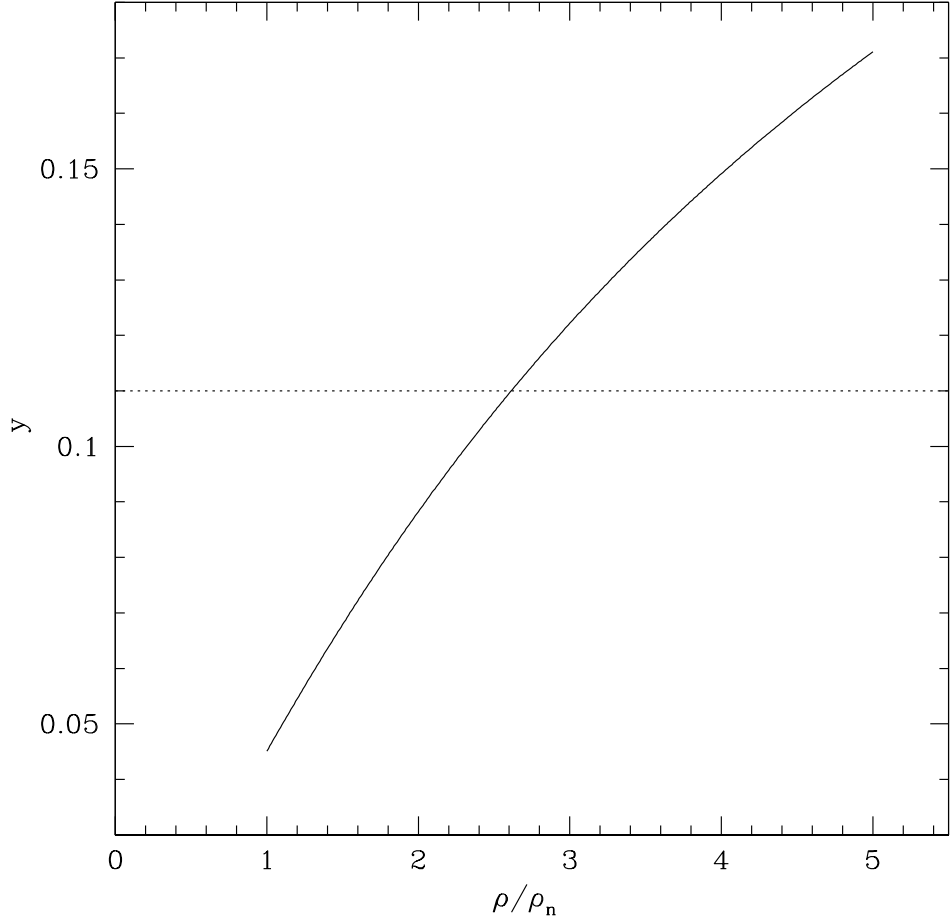}
\epsscale{1}
 \end{center}
\caption{\emph{Proton fraction $y$ versus baryon density in our model. Direct urca neutrino emission becomes allowed at $y\approx 0.11$ or $\rho\approx 2.5\rho_n$. 
\label{fig:yvsrho}
}}
\end{figure}

\beqy
\frac{1}{m_p^*}&=&\frac{1}{M}+\frac{2g_V\omega_0}{p_{F_p}^2}+\frac{g_{\rho}\rho_0}{p_{F_p}^2} \\
\frac{1}{m_n^*}&=&\frac{1}{M}+\frac{2g_V\omega_0}{p_{F_n}^2}-\frac{g_{\rho}\rho_0}{p_{F_n}^2}
\eeqy

where $p_{F_p}=(3\pi^2\rho_V y)^{1/3}\,;\quad p_{F_n}=(3\pi^2\rho_V (1-y))^{1/3}$. From Figure~\ref{fig:yvsrho} for the proton fraction $y(\rho_V)$, it appears that neutrino emission by the direct urca process sets in for densities close to $2.5\rho_n$. 
The modified urca, in which energy-momentum conservation is facilitated by a spectator nucleon, occurs at all densities, and the associated neutrino rate is (Friman \& Maxwell (1979))

\beqy
&&\epsilon_{\nu}^{\rm MUrca}=2.45\times 10^{21}\left(\frac{m_p^*}{\rm GeV}\right)\left(\frac{m_n^*}{\rm GeV}\right)^3\alpha_{\rm Murca}\left(\frac{\rho_V}{\rho_n}\right)^{1/3}y^{1/3}T_9^8~ {\rm erg~cm}^{-3}{\rm s}^{-1}\nonumber\,; \\
&&\alpha_{\rm Murca}=2\left(\frac{1}{1+0.173\left(\frac{\rho_n}{\rho_V}\right)^{2/3}\frac{y^{4/3}}{(y-y^2)^{2/3}}}\right)\quad . \label{murca}
\eeqy

\noindent The calculation is described in detail in Friman \& Maxwell
(1979), who worked in neutron matter with a phenomenological
description of nuclear forces, and explicit nucleon fields. In our
case, we used the Skyrme model to derive eqn.(\ref{murca}). The Skyrme model
can reproduce the long-range part of the $NN$ interaction, namely, the
one-pion exchange (Nyman \& Riska 1990), which is the most relevant piece for the modified urca process at these densities. In obtaining the above
result, eqn.(\ref{murca}), the following points should be noted. We
used the pion-nucleon coupling constant value $g_{\pi
  NN}=11.9$. This is the predicted value in the model (Adkins
  and Nappi 1984b) that we are using, compared to the actual
  experimental value of 13.2. The model parameters are chosen to
  reproduce the nucleon (938.9 MeV) and Delta (1232 MeV)
  mass. Although we have used a relativistic model for the nucleons
thus far, we approximated their dispersion relations by their
non-relativistic counterparts for calculational simplicity. Thus,
$E_p=p^2/2M - p^4/8M^3 +$ .. . The relative error in retaining only
the leading term is about $p_F^2/4M^2\sim 10\%$ if we consider
densities as high as $2\rho_n$. This does not affect our final result
for the emissivity in any considerable way. At higher densities, the
Fermi momentum of the nucleon is not small compared to its effective
mass $M$, and a fully relativistic treatment is required. Here, we
restrict our application to Skyrmion stars with central densities
$\rho_c\leq2\rho_n$ so that our non-relativistic approximations hold.
This restriction to low density also means that the direct urca
process remains forbidden. From the plots for mass versus central
density and mass versus radius, we observe that this limits us to
$M\leq 2M_{\odot}$ and $R\leq 15$km. Thus, we are well within current
observational limits.

\vskip 0.2cm

\noindent The neutrino luminosity $L_{\nu}$ is now determined as

\beq
L_{\nu}=4\pi\int_0^{R} dr~r^2~\epsilon_{\nu}(\rho(r))\,\,, \label{nulum}
\eeq
where $R$ denotes the radius of the star. The specific heat of the baryons can be expressed as

\beq
c_V=2\int\frac{d^3p}{(2\pi)^3}(\epsilon_p-\mu)\frac{\partial f(\epsilon_p)}{\partial T}\,\,
\eeq
where $f(\epsilon_p)=(1+{\rm exp}(\frac{\epsilon_p-\mu}{T}))^{-1}$ and we have omitted a term that is negligible in the degenerate limit. In this limit, one can extract the density of states $D(\epsilon_F)$ at the Fermi surface and express the specific heat as

\beq
c_V(\rho)=\frac{T}{3}\left[p_{F_n}\sqrt{p_{F_n}^2+M^2}+p_{F_p}\sqrt{p_{F_p}^2+M^2}\right]
\eeq

The total specific heat is then

\beq
C_V=4\pi\int_0^{R} dr~r^2~c_V(\rho(r))\,\,. \label{nuheat}
\eeq
With eqns.(\ref{nulum}) and (\ref{nuheat}), we can now proceed to compute the cooling curve for the star. In fact, an analytical estimate is easily made (Page et al. 2004). Performing the integrations in eqns.(\ref{nulum}) and (\ref{nuheat}), we obtain 

\beq
L_{\nu}=NT^8;\quad C_V=CT\,\,.
\eeq
where $C$ and $N$ depend on density but not on temperature. The cooling equation (\ref{cool}) can then be integrated to yield

\beq 
T=\left(\frac{C}{6N}\right)^{1/6}t^{-1/6}\label{Ttcool} 
\eeq 

for long-term cooling in the neutrino dominated epoch. The ``internal''
temperature $T$ is, strictly speaking, the temperature at the bottom
of the crust, and can be related to the temperature at the surface of
the crust via the so-called Tsuruta law (Tsuruta 1979) which is an
interpolation between the surface and the crust temperatures.  The
formula for thick crusts, as is the case in our model, is (see also
Shapiro \& Teukolsky 1983, p330) $T_{\rm b}= (10 T_{\rm m})^{2/3}$;
the subscript ``m" stands for mantle so that $T=T_{\rm m}$, the subscript
``b'' for the bottom of the stellar envelope (which is also the top of
the crust).  It is likely, however, that very soon after the neutron star's formation, the crust becomes superfluid, and is
therefore almost isothermal. We can then obtain the surface
temperature $T_e$ at the top of the stellar envelope (atmosphere)
directly by employing the $T_e-T_b$ relation, first derived in
Gudmundsson et al. (1982), and utilized for practical cooling
simulations in Page et al. (2004). With $T_b=T$, we find

\beq 
T_e=0.87\times 10^6 (g_{s_{14}})^{1/4}(T_{8})^{0.5+\alpha}\quad.
\eeq 

\noindent Here, $g_{s_{14}}$ is the surface gravity $g_s$ measured in $10^{14}$
 cgs units (see eqn.(\ref{gs14})),  $T_{8}$ is temperature in units of $10^8$K and $\alpha$ parameterizes the composition of the envelope (we choose $\alpha=0.05$ which is typical of an iron envelope).

\begin{figure}[htb]
\begin{center}
\epsscale{.8}
\plotone{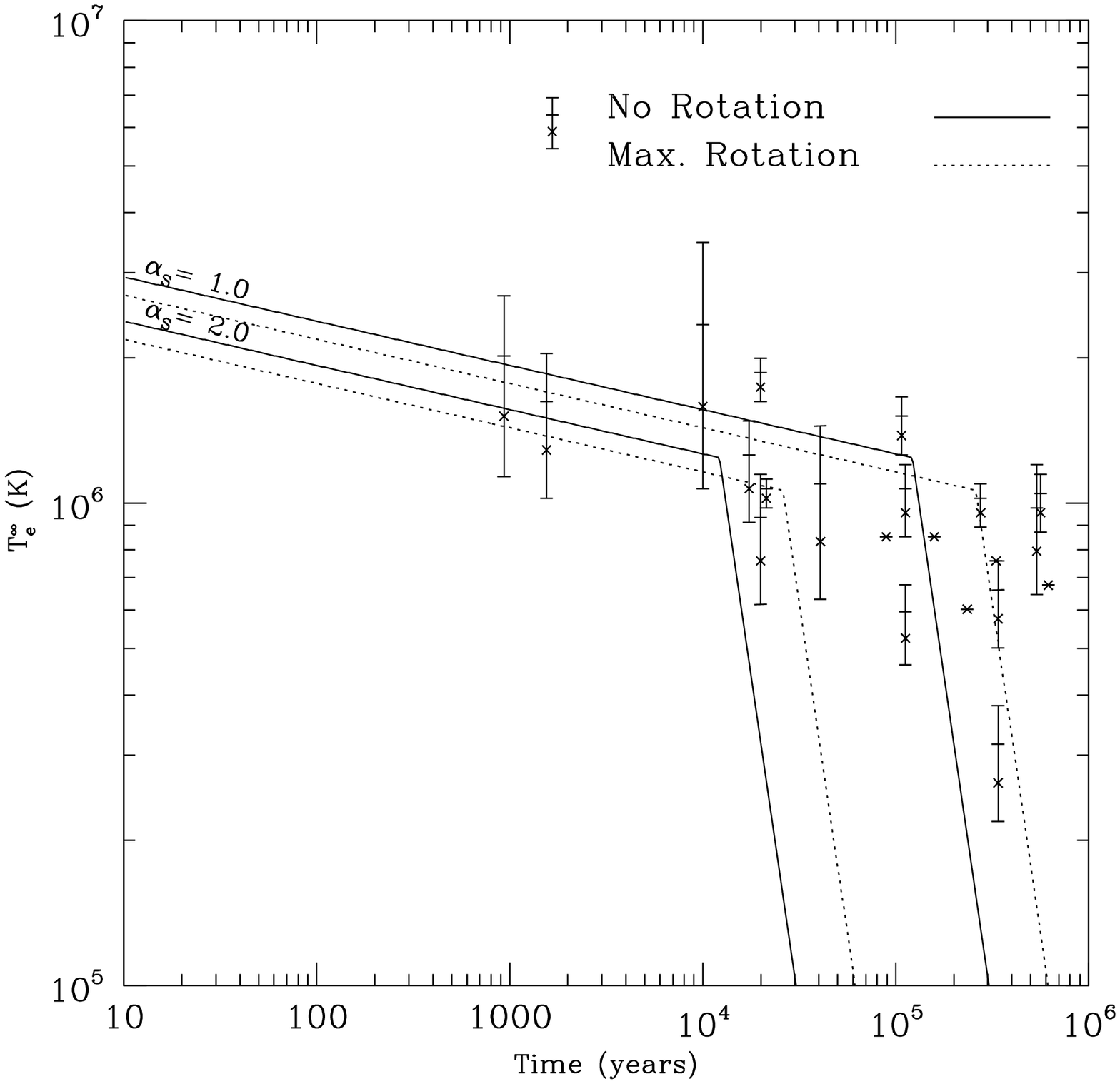}
\epsscale{1}
 \end{center}
\caption{\emph{Cooling curve for Skyrmion stars with central density $\rho_c=2\rho_n$, for maximally rotating (dashed) and static (solid) configurations. $\alpha_s$ is a parameter related to crustal superfluidity (see text for details). Temperature data points are from Schaab et al. 1999. 
}}
\label{fig:cooling}
\end{figure}

\vskip 0.2cm 
\noindent The effective temperature as seen by a distant observer is
then $T_{\rm e}^{\infty}=T_{\rm e}\sqrt{1-R_{\rm Sch.}/R}$ where
$R_{\rm Sch.}$ is the star's Schwarschild radius. The resulting
cooling curves ($T_{\rm e}^{\infty}$ versus time) are plotted in
Figure~\ref{fig:cooling} for Skyrmion stars with central density
$\rho_{\rm c}=2\rho_n$. The cooling curves for the non-rotating
configurations correspond to $M=2.2M_{\odot},R=16\ {\rm km}$ and those
for the maximally rotating ones correspond to $M=3.0M_{\odot},R=22\
{\rm km}$. Note that we included photon cooling which starts to
dominate when $L_{\gamma}\gg L_{\nu}$. In that era, the $T$ vs. $t$
behavior is 
\beq 
T=\left(\frac{C}{4\alpha
S}\right)^{1/4\alpha}t^{-\frac{1}{4\alpha}} \,\,, 
\eeq 
where $C$ and $\alpha$ are the same constants as before, and 
\beqy 
S&=&4.08\times
10^{16-32\alpha}R_{10}^2~g_{s_{14}}~{\rm ergs/sec~K}^{2+4\alpha}\,;\\
g_{s_{14}}&=&1.334\frac{M}{M_{\odot}}\frac{{\rm
e}^{\Lambda_s}}{R_{10}^2}\,;\label{gs14}\\ {\rm
e}^{\Lambda_s}&=&\frac{1}{\sqrt{1-0.296\frac{M}{M_{\odot}}\frac{1}{R_{10}}}}
\eeqy
with $R_{10}$ the stellar radius in units of 10 km. The photon
luminosity in units of ergs/sec is given by
$L_{\gamma}=ST^{2+4\alpha}$. Irrespective of the initial temperature
($T(t=0)$), we observe that $T_e^{\infty}$ in Figure~\ref{fig:cooling}
traces the temperature data, which are mostly constrained only by
upper limits. Photon cooling starts to dominate close to a million
years after the neutron star's formation. The exceedingly rapid
cooling of the star then makes it thermally invisible ($T_e^{\infty}$
below $10^5$K). The parameter $\alpha_s$ accounts for a possible
suppression of the nucleonic specific heat by superfluidity. The
electrons then control the specific heat, given by
$c_V^e=10^{-\alpha_s}c_V^{n,p({\rm ungapped})}$ with $\alpha_s\simeq2$
a typical value (Page et al., 2004). We have shown curves for two
cases, $\alpha_s=1.0,2.0$. Note also that for the rotating case,
neutrino cooling is more efficient but photon cooling takes over at a
later stage as compared to the non-rotating case. This is due to the
smaller crustal mass fraction for the rotating case. Our estimates of
the cooling rate are bound to be affected mildly by inclusion of other
neutrino emission processes such as neutrino bremsstrahlung and Cooper
pair-breaking and formation processes, as well as effects from
time-evolving stellar envelopes, but our results capture the essential
cooling history of a Skyrmion star.
\
\section{Conclusions}
\label{sec:conclusion}

In this work, we have examined the case for considering the Skyrme model as a viable description of dense nuclear and neutron-rich matter, with direct application to neutron star interiors. There are clear indications that conventional neutron matter equations of state cannot explain the observation of some very heavy neutron stars, unless the central densities are assumed to be fairly high. The Skyrme Lagrangian, within the mean-field approach employed here, and with the inclusion of the dilaton and vector mesons, serves to generate quite stiff equations of state that can describe the structure of these heavy stars. Importantly, the Skyrme model incorporates features of QCD such as scale breaking (the trace anomaly) and chiral fields that map to low-lying excitations of the vacuum. It should be noted, however, that the use of the dilaton as an interpolating field to mimic the (large) scale-breaking in QCD appears at odds with the estimated glueball mass of 1.5 GeV and its stiffness in matter (Birse (1994)). A better in-medium description of the Skyrme model would have to take this fact into consideration.

\vskip 0.2cm

\noindent The mass-radius curves for Skyrmion stars are distinct from ordinary neutron stars, essentially due to the qualitatively different nature of the dilaton potential, which respects the scale anomaly, and is fit to nuclear matter properties. We have described both rotating and non-rotating configurations for Skyrmion stars, the essential difference between them being the crustal mass fraction, and the total mass that they can support (rotating configurations can support considerably more mass). Their masses and radii lie in the range $0.4 \le M/M_{\odot} \le 3.6$ and $13~{\rm km} \le R \le 23~{\rm km}$, respectively.  The minimum spin period lies between $0.7~{\rm ms}\le P \le 2.1~{\rm ms}$. Skyrmion stars offer a possible explanation for observations of large mass neutron stars. Such stars are also indirectly hinted at in astrophysical models of the r-process, cooling curves of neutron stars, and have been examined in population synthesis studies. 

\vskip 0.2cm

\noindent We have also studied cooling of Skyrmion stars by
identifying the dominant neutrino emission process, the modified urca,
and computing the corresponding emissivity. The underprediction of
$g_A$ and $g_{\pi NN}$ in the model implies that the neutrino
emissivity is also underestimated by a factor$\sim 5$. However, this
has only a slight effect on the cooling rate since the temperature
depends on the 1/6th power of the ratio $C/N$ (see
eqn.(\ref{Ttcool})). Along with the specific heat of degenerate
matter, this allows us to sketch the cooling history of the star from
tens of years up to a million years when the star rapidly cools to
unobservably low temperatures. A more accurate cooling curve can be
produced by taking into account the thermal conductivity of the
material, density dependence of superfluid gaps, and other neutrino
emission processes. We have included photon cooling, assuming a
generic envelope at the surface. While these surface structures are
not direct consequences of our model, they are ``pinned on'' here for
a more realistic picture, and to enable qualitative comparisons to the
neutron star temperature data.
 
\vskip 0.2cm

\noindent An additional contribution to the neutrino flux that we have ignored in this model comes from the electroweak decay of charged pions to neutrinos. Furthermore, neutral pions can also decay into neutrino-antineutrino pairs. Although this process is forbidden by helicity conservation in vacuum, the presence of a medium implies that decay can occur in a frame moving with respect to the preferred rest frame. In fact, this process has been addressed within the Skyrme model very recently (Kalloniatis et al. 2005). The reason we have neglected such ``pionic'' contributions is that the corresponding neutrino rate is penalized by an exponential factor of ${\rm exp}(-m_{\pi}/T)$. Since $T\ll m_{\pi}$ in our case, the neutrino emissivity is exponentially small. The dominant contribution in our model is indeed from the modified urca process, and from the direct urca process at much higher densities.

\vskip 0.2cm

\noindent Assuming a  mean-field approach to the Skyrme fluid and ignoring Skyrmion overlap at high density is no doubt a simplification, but our aim in this work has been to gauge the model's main features, and examine its potential for the astrophysics of neutron stars. While improvements can surely be made on the theoretical aspects of the assumed model, we find that it displays qualitative features regarding its mass-radius curve that are very different from ordinary neutron stars. This could bear importantly upon the observations of heavier mass and/or large radius neutron stars as well as their cooling history. 

\begin{acknowledgements}
\noindent Acknowledgements: P.J. thanks the Department of Physics and
Astronomy at the University of Calgary, where this work was initiated,
for its hospitality. The authors are grateful to an anonymous referee
for suggestions improving the overall quality of the paper. The
research of R.O. is supported by grants from the Natural Science and
Engineering Research Council of Canada (NSERC) as well as the Alberta
Ingenuity Fund (AIF). P.J. is supported by the Department of Energy,
Office of Nuclear Physics, contract no. W-31-109-ENG-38.

\end{acknowledgements}

\end{document}